# Quantum Dipolar Magnon Drag in Conducting Magnetic Bilayers

Ling Gan[1*], and Shufeng Zhang[1†]

[1]Physics Department, University of Arizona, Tucson AZ
Ling Gan: ORCID 0009-0007-0263-7602

**ABSTRACT**. We predict electrical drag between two electrically isolated metallic ferromagnets mediated by quantum dipolar magnon coupling. A charge current in one layer generates a nonequilibrium magnon current whose momentum is transferred across an insulating spacer through interlayer dipolar interactions. The resulting magnon current in the passive layer produces a measurable voltage through electron–magnon scattering. We show the combination of resonant quantum momentum transfer and magnetic tunability through the control of magnetization configurations distinguish quantum dipolar drag from conventional drag mechanism.

Magnetic bilayers are fundamental building blocks of spintronic devices, exhibiting phenomena such as giant magnetoresistance, interlayer exchange coupling, and spin-transfer torque [1–8]. While interlayer magnetic dipolar interactions are conventionally treated as classical magnetostatic couplings [9,10], their quantum nature can mediate momentum transfer between magnons in spatially separated magnetic layers. Recent experiments have reported electrically induced angular momentum transfer between electrically isolated ferromagnets [11], raising the question of whether interlayer dipolar interactions can generate a measurable electrical drag effect.

Electrical drag has been extensively studied in spatially separated conductors through interlayer Coulomb interactions [12]. Magnon-mediated electrical drag has also been predicted and observed in heavy-metal/magnetic-insulator heterostructures [13–16], where magnons propagate within a common magnetic layer. Here, we propose a distinct mechanism: *quantum dipolar magnon drag between two electrically isolated metallic ferromagnets*, in which no direct electron transfer or magnon propagation across the insulating spacer is required.

As illustrated in Fig. 1, an applied charge current in the driving ferromagnet generates a nonequilibrium magnon current through electron–magnon scattering [17]. Quantum interlayer dipolar coupling transfers magnon momentum to the passive ferromagnet, where subsequent magnon–electron scattering produces an open-circuit drag voltage. We show that the dipolar interaction generates both magnon-conserving and anomalous pair processes, with the former exhibiting resonant enhancement when the magnon spectra of the two layers coincide. The resulting drag depends strongly on the magnetization orientations, providing a distinctive experimental signature and a means of controlling nonlocal electrical transport. We note that interlayer dipolar coupling has also been proposed for heat transfer between *magnetic insulators* separated by a non-magnetic heat insulator [18].

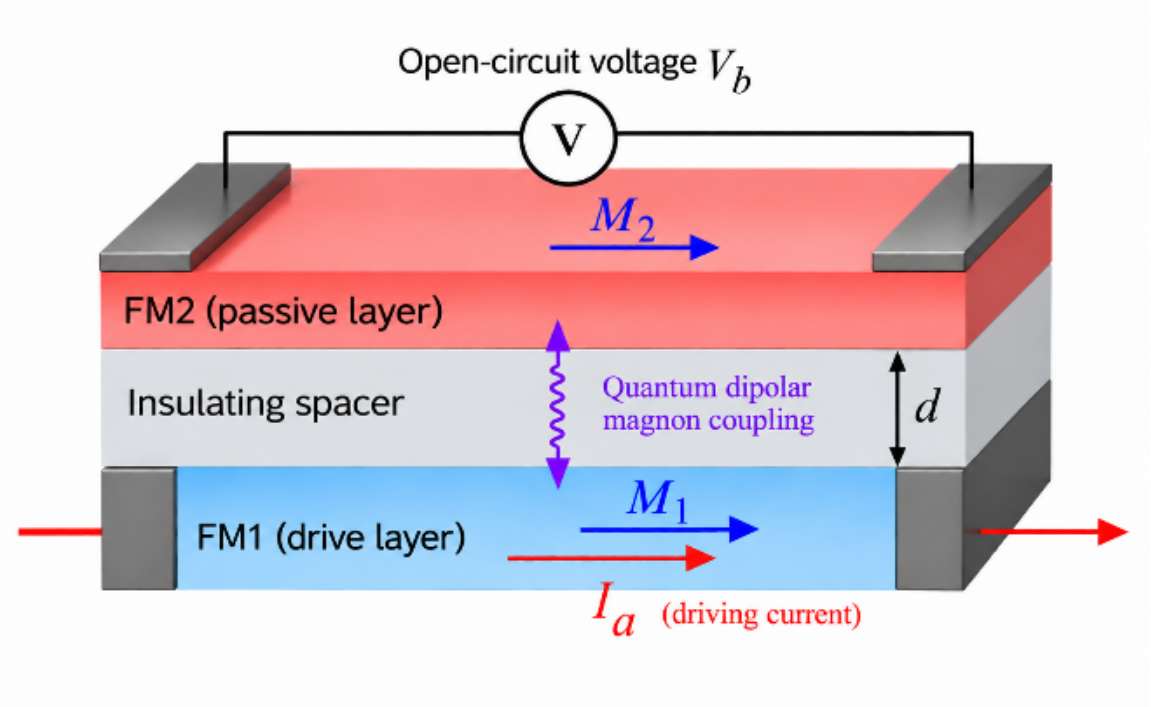

FIG 1. Schematic of quantum dipolar magnon drag between two electrically isolated metallic ferromagnets (FM1 and FM2) separated by an insulating spacer of thickness $d$. An applied charge current in FM1 generates a nonequilibrium magnon current, which transfers momentum to FM2 through quantum dipolar coupling and produces an open-circuit drag voltage $V_b$ in FM2.

Consider two thin magnetic layers, $M1$ and $M2$ (treated as 2D layers) separated by an insulating spacer of thickness $d$, shown in Fig. 1. In-plane coordinates are $\boldsymbol{\rho} = (x, y)$, and the displacement from a spin in layer $M2$ to a spin in layer $M1$ is $\mathbf{r} = (x, y, d)$. The interlayer dipolar interaction is written as

$$H_D = D_0 \sum_{ij} \sum_{\alpha\beta} S_{i\alpha}^a \, \mathcal{D}_{\alpha\beta}(\mathbf{r}_{ij}) \, S_{j\beta}^b, \qquad (1)$$

where $S_{i\alpha}^a$ and $S_{j\beta}^b$ are the $\alpha$ and $\beta$ components of the spins at sites $i$ and $j$ in layers $a$ $(M1)$ and $b$ $(M2)$, respectively, $D_0 = \mu_0 (g\mu_B)^2/4\pi$, and

$$\mathcal{D}_{\alpha\beta}(\mathbf{r}) = \frac{\delta_{\alpha\beta}}{|\mathbf{r}|^3} - \frac{3r_\alpha r_\beta}{|\mathbf{r}|^5}.$$

Retaining only terms bilinear in the magnon operators $a_\mathbf{k}$ and $b_\mathbf{k}$, the interlayer interaction Eq. (1), can be written as

$$H_D = H_1 + H_2, \tag{2}$$

where

$$H_1 = \sum_\mathbf{k} \left[\gamma_{1\mathbf{k}}\, a_\mathbf{k}^\dagger b_\mathbf{k} + \gamma_{1\mathbf{k}}^*\, a_\mathbf{k} b_\mathbf{k}^\dagger\right] \tag{3}$$

is the normal, magnon-conserving interlayer coupling, and

$$H_2 = \sum_\mathbf{k} \left[\gamma_{2\mathbf{k}}\, a_\mathbf{k}^\dagger b_{-\mathbf{k}}^\dagger + \gamma_{2\mathbf{k}}^*\, a_\mathbf{k} b_{-\mathbf{k}}\right] \tag{4}$$

is the anomalous pair-creation/pair-annihilation coupling. The momentum-dependent coefficients $\gamma_{1k}$ ($\gamma_{2k}$) depend on magnetization directions of the two layers. The explicit expressions for several magnetization configurations are derived in the Supplement Materials (SM) [19]. Throughout, the magnon operators in each layer are defined with respect to the local magnetization direction of that layer. This point is essential when the two magnetizations are antiparallel.

We now calculate the electrical drag generated by this quantum dipolar interaction, Eqs. (3) and (4). An applied electric current $j_e^a$ in the M1 layer is accompanied by a nonequilibrium magnon current $j_m^a/\mu_B = P_m j_e^a/e$ through electron-magnon scattering where the dimensionless constant $P_m$ for transition − metal ferromagnets was estimated to be about 0.2–0.5 at room temperature [17]. In the weak interlayer-dipolar coupling regime, the nonequilibrium magnon distribution in the drive layer $M1$ is fixed by the applied current and is unchanged, to leading order, by its coupling to layer $M2$. The induced distribution in layer $M2$ is therefore of order $|\gamma_{1k}|^2$ and $|\gamma_{2k}|^2$.

The magnon Hamiltonian is $H = H_0 + H_1 + H_2$, with the bare dispersions of the two layers,

$$H_0 = \sum_\mathbf{k} \left(\epsilon_{a\mathbf{k}} a_\mathbf{k}^\dagger a_\mathbf{k} + \epsilon_{b\mathbf{k}} b_\mathbf{k}^\dagger b_\mathbf{k}\right).$$

Let $\delta n_{a\mathbf{k}} = n_{a\mathbf{k}} - n_{a\mathbf{k}}^0$ denote the current-carrying magnon distribution in layer $M1$; its odd-in-momentum component determines the magnon current. To the lowest nonvanishing order in the interlayer dipolar coupling, the corresponding magnon distribution $n_{b\mathbf{k}} \equiv n_{b\mathbf{k}}^0 + \delta n_{b\mathbf{k}}$ in layer $M2$ obeys

$$\frac{d\delta n_{b\mathbf{k}}}{dt} = -\frac{\delta n_{b\mathbf{k}}}{\tau_{\mathrm{tr},b\mathbf{k}}} + (W_{1\mathbf{k}} - W_{2\mathbf{k}})\delta n_{a\mathbf{k}} + O(\gamma^4). \tag{5}$$

where

$$W_{1\mathbf{k}} = \frac{2|\gamma_{1\mathbf{k}}|^2}{\hbar^2} \frac{\Gamma_{1\mathbf{k}}}{\Gamma_{1\mathbf{k}}^2 + (\epsilon_{a\mathbf{k}} - \epsilon_{b\mathbf{k}})^2/\hbar^2} \tag{6}$$

and

$$W_{2\mathbf{k}} = \frac{2|\gamma_{2\mathbf{k}}|^2}{\hbar^2} \frac{\Gamma_{2\mathbf{k}}}{\Gamma_{2\mathbf{k}}^2 + (\epsilon_{a\mathbf{k}} + \epsilon_{b\mathbf{k}})^2/\hbar^2} \tag{7}$$

are the normal and anomalous dipolar transfer rates, respectively. The derivations of (5)-(7) are given in SM [19]. The transport time $\tau_{\mathrm{tr},b\mathbf{k}}$ characterizes relaxation of the current-carrying, odd-in-$k$ magnon distribution and can be strongly affected by electron-magnon scattering; estimates based on the corresponding resistive contribution give a picosecond scale [20-24]. By contrast, the long-wavelength coherence time $T_{\mathrm{ik}} = \Gamma_{i\mathbf{k}}^{-1}$ is related to the magnetic-resonance linewidth and can be much longer in low-damping ferromagnets. We may take $T_{\mathbf{ik}} = 10^{-8}$ s as a representative long-wavelength value [24,25], while emphasizing that the finite-$k$ coherence relevant to the drag may differ and should ultimately be determined experimentally or microscopically.

In the steady state, Eq. (5) gives

$$\delta n_{b\mathbf{k}} = \tau_{\mathrm{tr},b\mathbf{k}}(W_{1\mathbf{k}} - W_{2\mathbf{k}})\delta n_{a\mathbf{k}} \tag{8}$$

The induced magnon current is therefore

$$\mathbf{j}_b^m = \int \frac{d^2k}{(2\pi)^2} \mathbf{v}_\mathbf{k} \tau_{\mathrm{tr},b\mathbf{k}} (W_{1\mathbf{k}} - W_{2\mathbf{k}})\delta n_{a\mathbf{k}} \tag{9}$$

where $\epsilon_{b\mathbf{k}} = \Delta_0 + Dk^2$, and $\hbar\mathbf{v}_\mathbf{k} = 2\mathrm{D}\mathbf{k}$, with $\Delta_0$ and $D$ the magnon gap and stiffness. Several features of Eq. (9) are noteworthy. The normal process couples $a_\mathbf{k}$ to $b_\mathbf{k}$, whereas the anomalous process couples $a_\mathbf{k}$ to $b_{-\mathbf{k}}$. Both conserve the total momentum of the coupled two-layer system. However, after relabeling $-\mathbf{k} \to \mathbf{k}$ and using the odd parity of the current-carrying distribution, the anomalous channel contributes with the opposite sign to the induced magnon current in layer $M2$, producing the combination $W_{1\mathbf{k}} - W_{2\mathbf{k}}$ in Eqs. (5)–(9), see SM [19].

A second important distinction is energy. The normal channel becomes resonant when $\epsilon_{a\mathbf{k}} = \epsilon_{b\mathbf{k}}$ for the identical layers, $W_{1\mathbf{k}} = 2|\gamma_{1\mathbf{k}}|^2/(\hbar^2\Gamma_{1\mathbf{k}})$.

*Contact author: linggan@ariona.edu
†Contact author: zhangshu@arizona.edu

The anomalous pair channel instead contains the energy sum $\epsilon_{a\mathbf{k}} + \epsilon_{b\mathbf{k}}$ and is therefore generally off resonance. Consequently, when $\hbar\Gamma_{\boldsymbol{k}} \ll \epsilon_{\mathbf{k}}$, one has $W_{1\boldsymbol{k}} \gg W_{2\boldsymbol{k}}$ even when $|\gamma_{1\boldsymbol{k}}|$ and $|\gamma_{2\boldsymbol{k}}|$ are comparable.

The electrical response in layer $M2$ follows from momentum transfer from the induced magnons to the conduction electrons. We use the standard electron-magnon interaction,

$$H_{em} = \sum_{\mathbf{p},\mathbf{q}} M_{\mathbf{q}}\, c^{\dagger}_{\mathbf{p}+\mathbf{q}\downarrow} c_{\mathbf{p}\uparrow} b_q + h.c,$$

where $c_p^{\dagger}(c_p)$ is electron creation (annihilation) operator, and $M_{\mathbf{q}}$ is the $s$ - $d$ exchange coupling constant, that gives, to linear order in $\delta n_{b\mathbf{q}}$, the rate of momentum transfer to the electron system,

$$\frac{dP_e}{dt} = \int_{\mathbf{q}} d^2q\, \frac{\hbar\mathbf{q}}{\tau_{me}(\mathbf{q})}\delta n_{b\mathbf{q}}$$

where we have defined

$$\frac{1}{\tau_{me}(\mathbf{q})} = \frac{2\pi}{\hbar}\left|M_{\mathbf{q}}\right|^2 \int_{\mathbf{p}} d^2\mathbf{p}\, \left(f^0_{\mathbf{p}\uparrow} - f^0_{\mathbf{p}+\mathbf{q}\downarrow}\right) \delta\left(\epsilon^e_{\mathbf{p}+\mathbf{q}\downarrow} - \epsilon^e_{\mathbf{p}\uparrow} - \hbar\omega_{\mathbf{q}}\right). \tag{10}$$

In an open-circuit measurement the total electrical current in layer $M2$ vanishes. The electric field balances the magnon-induced force:

$$\mathbf{E}_b = \frac{1}{n_b e}\int d^2q \frac{\hbar\mathbf{q}\,\tau_{\mathrm{tr},b\mathbf{q}}}{\tau_{me}(\mathbf{q})}\left(W_{1\mathbf{q}} - W_{2\mathbf{q}}\right)\delta n_{a\mathbf{q}} \tag{11}$$

It remains to specify the non-equilibrium magnon density $\delta n_{ak}$ in the driving layer $M1$ to evaluate the electric field from Eq. (11). Recall from [17] that $\delta n_{aq} = n_{aq} - n^0_{aq} = \frac{\partial n^0_{aq}}{\partial \epsilon_{aq}} g_{aq}$ where $n^0_{aq} = \left(\exp(\beta\epsilon_{aq}) - 1\right)^{-1}$ is the equilibrium magnon density and $g_{aq}$ can be expanded in terms of Taylor series in the momentum, $g_{\boldsymbol{ap}} = p_x(A_0 + A_1\boldsymbol{p} + A_2\boldsymbol{p}^2 \ldots)$.

Next, we should estimate the magnitude and the separation dependence of the drag effect by assuming momentum independent $\tau_{tr,bk}$ and $T_{ik}$,. By taking $g_{\boldsymbol{ap}} = A_0 p_x$, we have

$$\frac{j^m_b}{j^m_a} = \frac{\int \frac{d^2k}{(2\pi)^2} \mathbf{v}_{\mathbf{k}}\tau_{\mathrm{tr},b\mathbf{k}}\, W_{1\mathbf{k}}\, \delta n_{a\mathbf{k}}}{\int \frac{d^2k}{(2\pi)^2} \mathbf{v}_{\mathbf{k}}\, \delta n_{a\mathbf{k}}} \sim \frac{S\,\tau_{\mathrm{tr},b\mathbf{k}} T_2}{\hbar^2}\left[\frac{\mu_0 (g\mu_B)^2 S}{4a_0^3}\right]^2 \left(\frac{a_0}{d}\right)^2 \ln^{-1}\left(\frac{k_B T}{\Delta_0}\right) \tag{12}$$

where we have used the high temperature approximation, $k_B T \gg \hbar\omega\left(k = \frac{1}{d}\right)$ such that the equilibrium magnon density $n^0(\omega_k) \simeq k_B T/(\hbar\omega_k)$ in calculating the numerate of the above equation, while the denominate is controlled by the ratio of the temperature and the magnon gap.

By placing $\tau_{\mathrm{tr},b\mathbf{k}} = 10^{-12}$ s, $T_{\mathrm{ik}} = 10^{-8}$ s, $d = 10$ nm, and $k_b T = 100\Delta_0$ into Eq.(12) , we have $j^m_b/j^m_a \sim 10^{-4}$. The measurable open-circuit field should then be evaluated directly from Eq. (11), but it involves further parameters. Since a rough estimate of the ratio between the magnon current and the driving electric current in the layer $M1$ is about 0.2-0.5 [19] and we would expect that the inverse conversion rate in the passive layer $M2$ is similarly valued. Thus, in the open circuit measurement, the drag resistivity $\rho_d \equiv E_b/j_a$ $\sim 10^{-5}\rho_e$where $\rho_e$ is the normal resistivity of the layers.

The dependence of the magnon drag on the separation of the two layers is also show in Eq.~(12). The $1/d^2$ is derived from the high temperature approximation, $k_B T \gg \hbar\omega\left(k = \frac{1}{d}\right)$, in addition to the momentum independent relaxation times. At low temperature, the equilibrium density rapidly decays and only small $k$ magnons contribute to the magnon drag so that the factor $e^{-2kd}$ approaches unity and the leading spacer dependence becomes weak. These scaling forms are intended as limiting results; realistic momentum-dependent relaxation and dephasing will modify the crossover.

The most distinctive signatures of quantum dipolar drag is its dependence on magnetization orientation. The coefficients $\gamma_{1\mathbf{k}}$ and $\gamma_{2\mathbf{k}}$ depend on both the propagation direction and the local quantization axes. As derived in the SM [19], when the two magnetizations are switched from parallel to antiparallel, $\gamma_{1\mathbf{k}}$ and $\gamma_{2\mathbf{k}}$ are exchanged. In the case of perpendicular magnetization, $|\gamma_{1\mathbf{k}}|^2 = |\gamma_{2\mathbf{k}}|^2$, the parallel and antiparallel configurations have the same drag current. In the case of in-plane magnetization, $\gamma_{1\mathbf{k}}$ and $\gamma_{2\mathbf{k}}$ have a strong angular dependence. Averaging over the angular dependence over $|\gamma_{1\mathbf{k}}|^2$, we show, in Table I, the drag magnitudes for six magnetization configurations: parallel and antiparallel magnetization of $M1$ and $M2$ for magnetization in *x-axis*, *y-axis, and z-axis*. Noticeably, the antiparallel and parallel magnetization for the in-plane magnetization could differ by up to 49 times (the last two columns of Table 1). These configuration-dependent signatures provide a direct experimental test of the quantum dipolar mechanism.

*Contact author: linggan@ariona.edu
†Contact author: zhangshu@arizona.edu

TABLE I. The electric field in layer M2 induced by the magnon drag for six magnetization configurations, normalized by the value for $\hat{\mathbf{M}}_1 = \hat{\mathbf{M}}_2 = \hat{\mathbf{z}}$. The electric current is applied in the $+x$ direction. OOP (IP) denotes out-of-plane (in-plane) magnetization, while P (AP) refers to parallel (antiparallel) alignment of the magnetizations of two ferromagnetic layers.

| | |
|---|---|
| OOP and P: $\hat{\mathbf{M}}_1 = \hat{\mathbf{M}}_2 = \hat{\mathbf{z}}$ | 1 |
| OOP and AP: $\hat{\mathbf{M}}_1 = -\hat{\mathbf{M}}_2 = \hat{\mathbf{z}}$ | 1 |
| IP and P: $\hat{\mathbf{M}}_1 = \hat{\mathbf{M}}_2 = \hat{\mathbf{x}}$ | 5/8 |
| IP and AP: $\hat{\mathbf{M}}_1 = -\hat{\mathbf{M}}_2 = \hat{\mathbf{x}}$ | 21/8 |
| IP and P: $\hat{\mathbf{M}}_1 = \hat{\mathbf{M}}_2 = \hat{\mathbf{y}}$ | 1/8 |
| IP and AP: $\hat{\mathbf{M}}_1 = -\hat{\mathbf{M}}_2 = \hat{\mathbf{y}}$ | 49/8 |

Finally, it is instructive to compare quantum dipolar drag ) with the well-established Coulomb drag (CD) between spatially separated two-dimensional electron gases in which the drag resistivity scales as $\left(\frac{T}{T_F}\right)^2 (k_F d)^{-4}$ where $T_F$ and $k_F$ are the Fermi temperature and Fermi wave number [12]. Conventional CD originates from interlayer electron-electron Coulomb scattering and is reduced by screening and layer separation, whereas QDD is mediated by interlayer magnon-magnon coupling and is resonantly enhanced when the magnon spectra of the two layers are matched. Using representative metallic-ferromagnet parameters, the estimates above indicate that QDD can reach a readily measurable regime at room temperature, where the drag resistivity is about $\rho_D = 10^{-5}\rho_e$ and can exceed typical Coulomb-drag signals by orders of magnitude. In addition, QDD possesses an intrinsic magnetic control parameter absent in ordinary Coulomb drag: changing the magnetization directions modifies the normal and anomalous dipolar matrix elements and therefore the magnitude of the drag. The combination of resonant quantum momentum transfer, electrical detection, and magnetic tunability distinguishes QDD from conventional drag mechanisms [13] and provides a route to controllable nonlocal transport between electrically disconnected magnetic conductors.

This work is supported by NSF-ECCS-2401267.

*Contact author: linggan@ariona.edu

†Contact author: zhangshu@arizona.edu

*Contact author: linggan@ariona.edu

†Contact author: zhangshu@arizona.edu

# Supplement Materials: Quantum Dipolar Magnon Drag in Conducting Magnetic Bilayers

## I. Hamiltonian of Quantum Dipolar Coupling

### 1. Geometry and Dipolar Hamiltonian

Consider two 2D magnetic layers, $M1$ and $M2$, separated by an insulating spacer of thickness $d$. The in-plane coordinates are $\boldsymbol{\rho} = (x, y)$, and the displacement from a spin in layer $M2$ to a spin in layer $M1$ is

$$\mathbf{r} = (x, y, d), \qquad r = \sqrt{x^2 + y^2 + d^2}.$$

The interlayer dipolar interaction is,

$$H_D = D_0 \sum_{ij} \sum_{\alpha\beta} S_{i\alpha}^{a}\, \mathcal{D}_{\alpha\beta}(\mathbf{r}_{ij})\, S_{j\beta}^{b}, \qquad (1)$$

where $D_0 = \mu_0 (g\mu_B)^2/4\pi$, and

$$\mathcal{D}_{\alpha\beta}(\mathbf{r}) = \frac{\delta_{\alpha\beta}}{|\mathbf{r}|^3} - \frac{3\mathbf{r}_\alpha \mathbf{r}_\beta}{|\mathbf{r}|^5} = -\partial_\alpha \partial_\beta \frac{1}{r} \qquad (2)$$

### 2. Useful two-dimensional Fourier transforms of the dipolar tensor

By using the following identity,

$$F(\mathbf{k}, d) = \int d^2\rho \, \frac{e^{-i\mathbf{k}\cdot\boldsymbol{\rho}}}{\sqrt{\rho^2 + d^2}} = \frac{2\pi}{k} e^{-kd}, \qquad k = \sqrt{k_x^2 + k_y^2},$$

the Fourier transformation of the dipolar tensor $\widetilde{\mathcal{D}}_{\alpha\beta} \equiv \int d^2\rho \, \mathcal{D}_{\alpha\beta}(\mathbf{r})\, e^{-i\boldsymbol{k}\cdot\boldsymbol{\rho}}$ can be readily obtained,

$$\widetilde{\mathcal{D}}_{xx} = 2\pi e^{-kd} \frac{k_x^2}{k},$$

$$\widetilde{\mathcal{D}}_{yy} = 2\pi e^{-kd} \frac{k_y^2}{k},$$

$$\widetilde{\mathcal{D}}_{xy} = 2\pi e^{-kd} \frac{k_x k_y}{k},$$

$$\widetilde{\mathcal{D}}_{xz} = i\, 2\pi k_x e^{-kd},$$

$$\widetilde{\mathcal{D}}_{yz} = i\, 2\pi k_y e^{-kd},$$

and

$$\widetilde{\mathcal{D}}_{zz} = -2\pi k e^{-kd}$$

It is convenient to define the common factor of the above dipolar tensor matrix for later use,

$$\gamma_0(k) \equiv \frac{D_0 S}{2a_0^2}\left(2\pi k e^{-kd}\right) \qquad (3)$$

where $a_0^2$ is the area per spin for a square lattice.

**3. Quantized dipolar Hamiltonians**

The Holstein-Primakoff transformation chooses the quantization axis parallel to the magnetization direction. In the following we will separately determine the quantum dipolar Hamiltonians in four cases: M1 and M2 are parallel or antiparallel, and they are either perpendicular to the plane the layers or in the plane of the layers.

### 3a. $\mathbf{M1} \parallel \mathbf{M2} \parallel +\hat{\mathbf{z}}$

The quantization axis is $+\hat{z}$. To lowest order in the Holstein-Primakoff expansion for the layer M1,

$$S^z = \left(S - a^\dagger a\right),$$

$$S^x = \sqrt{\frac{S}{2}}\left(a + a^\dagger\right),$$

$$S^y = -i\sqrt{\frac{S}{2}}\left(a - a^\dagger\right),$$

and similarly for the layer M2 with the boson operator denoted as $b$.

The transverse part of the interlayer dipolar Hamiltonian is

$$\mathrm{H_D} = D_0 \sum_{ij} \left[\mathcal{D}_{xx} S_{ix}^a S_{jx}^b + \mathcal{D}_{yy} S_{iy}^a S_{jy}^b + \mathcal{D}_{xy}\left(S_{ix}^a S_{jy}^b + S_{iy}^a S_{jx}^b\right)\right].$$

After inserting the magnon operators,

$$H_D = \frac{D_0 S}{2} \sum_{ij} \left\{\left(\mathcal{D}_{xx} + \mathcal{D}_{yy}\right)\left(a_i^\dagger b_j + b_j^\dagger a_i\right) + \left(\mathcal{D}_{xx} - \mathcal{D}_{yy} + 2i\mathcal{D}_{xy}\right) a_i^\dagger b_j^\dagger + \mathrm{h.c.}\right\}.$$

Replacing the boson operators by their Fourier transformation, $a_i = \sum_k a_k e^{ik\cdot R_i}$ and $b_i = \sum_k b_k e^{ik\cdot R_i}$ we have,

$$\mathrm{H_D} = H_1 + H_2, \qquad (3)$$

where

$$H_1 = \sum_{\mathbf{k}} \left[\gamma_{1\mathbf{k}}\, a_{\mathbf{k}}^\dagger b_{\mathbf{k}} + \gamma_{1\mathbf{k}}^*\, a_{\mathbf{k}} b_{\mathbf{k}}^\dagger\right] \qquad (4)$$

is the normal, magnon-conserving interlayer coupling, and

$$H_2 = \sum_{\mathbf{k}} \left[\gamma_{2\mathbf{k}}\, a_{\mathbf{k}}^\dagger b_{-\mathbf{k}}^\dagger + \gamma_{2\mathbf{k}}^*\, a_{\mathbf{k}} b_{-\mathbf{k}}\right] \qquad (5)$$

is the anomalous pair-creation/pair-annihilation coupling, with the coefficients (in the continuum limit of lattice)

$$\gamma_{1\mathrm{k}} = \tfrac{D_0 S}{2} \sum_{\boldsymbol{\rho}} e^{-i\mathbf{k}\cdot\boldsymbol{\rho}} \left(\mathcal{D}_{xx} + \mathcal{D}_{yy}\right) = \gamma_0(k) \qquad (6)$$

and

$$\gamma_{2k} = \tfrac{D_0 S}{2} \sum_{\boldsymbol{\rho}} e^{-i\mathbf{k}\cdot\boldsymbol{\rho}} \left(\mathcal{D}_{xx} - \mathcal{D}_{yy} + 2i\mathcal{D}_{xy}\right) = \gamma_0(k) e^{2i\phi} \qquad (7)$$

where we have defined $k_x = k\cos\phi$, and $k_y = k\sin\phi$, and used the relations in the previous section 2.

### 3b. $\mathbf{M1} = +M\hat{\mathbf{z}}$, and $\mathbf{M2} = -M\hat{\mathbf{z}}$

The quantization axis for the M2 layer is now $-\hat{\mathbf{z}}$. Choose a right-handed local frame

$$\hat{\mathbf{x}}_b{}' = \hat{\mathbf{x}}, \qquad \hat{\mathbf{y}}_b{}' = -\hat{\mathbf{y}}, \qquad \hat{\mathbf{z}}_b{}' = -\hat{\mathbf{z}}.$$

Then, in laboratory components,

$$S_b^x = \sqrt{\frac{S}{2}}\left(b + b^\dagger\right),$$

$$S_b^y = +i\sqrt{\frac{S}{2}}\left(b - b^\dagger\right),$$

whereas layer $a$ still has

$$S_a^y = -i\sqrt{\frac{S}{2}}\left(a - a^\dagger\right).$$

Substitution into the same dipolar Hamiltonian gives

$$H_D = \frac{D_0 S}{2}\sum_{ij}\left\{\left(\mathcal{D}_{xx} - \mathcal{D}_{yy} + 2i\mathcal{D}_{xy}\right)a_i^\dagger b_j + \left(\mathcal{D}_{xx} + \mathcal{D}_{yy}\right)a_i^\dagger b_j^\dagger + \mathrm{h.c.}\right\}.$$

Thus, the Hamiltonian has the same forms of Eq. (3)-(5) with the normal and anomalous coefficients exchanged, i.e.,

$$\gamma_{1k} = \gamma_0(k)e^{2i\phi} \qquad (8)$$

$$\gamma_{2\mathrm{k}} = \gamma_0(k) \qquad (9)$$

### 3c. $\mathbf{M}_a \parallel \mathbf{M}_b \parallel +\hat{\mathbf{x}}$

Take the current direction and both magnetizations along $x$. The quantization axis is now $x$,

$$S^x = \left(S - a^\dagger a\right),$$

$$S^y = \sqrt{\frac{S}{2}}\left(a + a^\dagger\right),$$

$$S^z = -i\sqrt{\frac{S}{2}}\left(a - a^\dagger\right).$$

The transverse dipolar interaction lies in the $y - z$ sector:

$$H_D = D_0\sum_{ij}\left[\mathcal{D}_{yy}S_{iy}^a S_{jy}^b + \mathcal{D}_{zz}S_{iz}^a S_{jz}^b + \mathcal{D}_{yz}\left(S_{iy}^a S_{jz}^b + S_{iz}^a S_{jy}^b\right)\right].$$

After the Holstein-Primakoff substitution,

$$H_D = \frac{D_0 S}{2}\sum_{ij}\{(\mathcal{D}_{yy}+\mathcal{D}_{zz})(a_i^\dagger b_j + b_j^\dagger a_i) + (\mathcal{D}_{yy}-\mathcal{D}_{zz}+2i\mathcal{D}_{yz})a_i^\dagger b_j^\dagger + \text{h.c.}\}.$$

The real-space tensor combinations are

$$\mathcal{D}_{yy}+\mathcal{D}_{zz} = \frac{2x^2-y^2-d^2}{(x^2+y^2+d^2)^{5/2}},$$

and

$$\mathcal{D}_{yy}-\mathcal{D}_{zz}+2i\mathcal{D}_{yz} = -\frac{3(y+id)^2}{(x^2+y^2+d^2)^{5/2}}.$$

Thus, the Hamiltonians remain same as Eq.(3)-(5) with the coefficients,

$$\gamma_{1k} = \frac{D_0 S}{2}\sum_{\boldsymbol{\rho}} e^{-i\mathbf{k}\cdot\boldsymbol{\rho}}\left(\mathcal{D}_{yy}+\mathcal{D}_{zz}\right) = -\gamma_0(k)\cos^2\phi, \qquad (10)$$

and

$$\gamma_{2\mathrm{k}} = \frac{D_0 S}{2}\sum_{\boldsymbol{\rho}} e^{-i\mathbf{k}\cdot\boldsymbol{\rho}}\left(\mathcal{D}_{yy}-\mathcal{D}_{zz}+2i\mathcal{D}_{yz}\right) = \gamma_0(k)(1-\sin\phi)^2. \qquad (11)$$

Note that these coefficients are strongly dependent on the direction of magnon propagation.

### 3d. Antiparallel in-plane magnetizations: $\mathbf{M}_a = +M\hat{\mathbf{x}}, \mathbf{M}_b = -M\hat{\mathbf{x}}$

Choose a right-handed local frame for layer $b$,

$$\hat{\mathbf{x}}_b{}' = -\hat{\mathbf{x}}, \qquad \hat{\mathbf{y}}_b{}' = \hat{\mathbf{y}}, \qquad \hat{\mathbf{z}}_b{}' = -\hat{\mathbf{z}}.$$

Then

$$S_b^y = \sqrt{\frac{S}{2}}(b+b^\dagger),$$

but

$$S_b^z = +i\sqrt{\frac{S}{2}}(b-b^\dagger),$$

whereas layer $a$ has the opposite sign in $S_a^z$.

Substituting into the dipolar interaction yields

$$H_{D,\parallel}^{AP} = \frac{D_0 S}{2}\sum_{ij}\{(\mathcal{D}_{yy}-\mathcal{D}_{zz}+2i\mathcal{D}_{yz})a_i^\dagger b_j + (\mathcal{D}_{yy}+\mathcal{D}_{zz})a_i^\dagger b_j^\dagger + \text{h.c.}\}.$$

After the substitution of the magnon operators by their Fourier transform, we find, once again, the normal and abnormal coefficients are exchanged compared to the parallel alignment of the magnetization, namely,

$$\gamma_{1k} = \gamma_0(k)(1-\sin\phi)^2 \qquad (12)$$

And

$$\gamma_{2k} = -\gamma_0(k)\cos^2\phi \qquad (13)$$

### 4. Summary of the four configurations

Taken together from Eqs.(3)-(13), the quantum dipolar Hamiltonian between two magnetic layers is

$$\mathrm{H_D} = H_1 + H_2,$$

where

$$H_1 = \sum_{\mathbf{k}}\left[\gamma_{1\mathbf{k}}\, a_{\mathbf{k}}^{\dagger} b_{\mathbf{k}} + \gamma_{1\mathbf{k}}^{*}\, a_{\mathbf{k}} b_{\mathbf{k}}^{\dagger}\right]$$

is the normal, magnon-conserving interlayer coupling, and

$$H_2 = \sum_{\mathbf{k}}\left[\gamma_{2\mathbf{k}}\, a_{\mathbf{k}}^{\dagger} b_{-\mathbf{k}}^{\dagger} + \gamma_{2\mathbf{k}}^{*}\, a_{\mathbf{k}} b_{-\mathbf{k}}\right]$$

is the anomalous pair-creation/pair-annihilation coupling. The coefficients for the four configurations are

| **Table S1: Magnetic configuration dependence** | $\gamma_{1\mathbf{k}}$ | $\gamma_{2\mathbf{k}}$ |
|---|---|---|
| Perpendicular, P: $+\hat{z}, +\hat{z}$ | $\gamma_0(k)$ | $\gamma_0(k)e^{2i\phi}$ |
| Perpendicular, AP: $+\hat{z}, -\hat{z}$ | $\gamma_0(k)e^{2i\phi}$ | $\gamma_0(k)$ |
| In-plane, P: $+\hat{x}, +\hat{x}$ | $-\gamma_0(k)\cos^2\phi$ | $\gamma_0(k)(1-\sin\phi)^2$ |
| In-plane, AP: $+\hat{x}, -\hat{x}$ | $\gamma_0(k)(1-\sin\phi)^2$ | $-\gamma_0(k)\cos^2\phi$ |

The common scale is $\gamma_0(k) \equiv \frac{D_0 S}{2a_0^2}\left(2\pi k e^{-kd}\right)$

## II. Magnon momentum transfer rates by quantum dipolar coupling

The magnon Hamiltonian of the two layers with the dipolar interaction is,

$$H = H_0 + H_1 + H_2 = \sum_{\boldsymbol{k}} \omega_{a\boldsymbol{k}} a_{\boldsymbol{k}}^{+} a_{\boldsymbol{k}} + \omega_{b\boldsymbol{k}} b_{\boldsymbol{k}}^{+} b_{\boldsymbol{k}} + \sum_{\boldsymbol{k}} (\gamma_{1\boldsymbol{k}} a_{\boldsymbol{k}}^{+} b_{\boldsymbol{k}} + \gamma_{1\boldsymbol{k}}^{*} a_{\boldsymbol{k}} b_{\boldsymbol{k}}^{+}) + (\gamma_{2\boldsymbol{k}} a_{\boldsymbol{k}}^{+} b_{-\boldsymbol{k}}^{+} + \gamma_{2\boldsymbol{k}}^{*} a_{\boldsymbol{k}} b_{-\boldsymbol{k}})$$

We consider that the magnon distribution in M1, $n_{a\mathbf{k}} = n_{a\boldsymbol{k}}^0 + \delta n_{a\mathbf{k}}$, is given by the applied electric current in M1 and is not affected by the dipolar coupling. We determine $n_{\mathrm{b}\mathbf{k}}$ in the weak coupling limit.

### 1. Normal dipolar term

Introduce the quantum coherence between the magnons in M1 and M2,

$$P_{\boldsymbol{k}} = < a_{\boldsymbol{k}}^{+} b_{\boldsymbol{k}} >$$

which satisfies the equation of motion,

$$\frac{dP_{\boldsymbol{k}}}{dt} = \frac{i}{\hbar} < [P_{\boldsymbol{k}}, H_0 + H_1] > -\Gamma_{1\boldsymbol{k}} P_{\boldsymbol{k}} = \frac{i}{\hbar}(\epsilon_{a\boldsymbol{k}} - \epsilon_{b\boldsymbol{k}}) P_{\boldsymbol{k}} - \frac{i\gamma_{1\boldsymbol{k}}^{*}}{\hbar}(\delta n_{a\boldsymbol{k}} - \delta n_{b\boldsymbol{k}}) \text{ -}\Gamma_{1\boldsymbol{k}} P_{\boldsymbol{k}}$$

Where we introduced a dephasing parameter $\Gamma_{1\boldsymbol{k}}$. In the steady state, $\frac{dP_{\boldsymbol{k}}}{dt} = 0$, we have,

$$P_{\boldsymbol{k}} = -\frac{i\gamma_{1\boldsymbol{k}}^{*}}{\hbar} \frac{\delta n_{a\boldsymbol{k}} - n_{b\boldsymbol{k}}}{\Gamma_{1\boldsymbol{k}} - i(\epsilon_{a\boldsymbol{k}} - \epsilon_{b\boldsymbol{k}})/\hbar}$$

The above expression is related to the transfer rate, because

$\frac{dn_{b\boldsymbol{k}}}{dt} = \frac{i}{\hbar} < [b_{\boldsymbol{k}}^{+} b_{\boldsymbol{k}}, H_1] >= -\frac{2}{\hbar} Im(\gamma_{1\boldsymbol{k}} P_{\boldsymbol{k}}) \equiv W_{1\boldsymbol{k}}(\delta n_{a\boldsymbol{k}} - \delta n_{b\boldsymbol{k}}) = W_{1\boldsymbol{k}} \delta n_{a\boldsymbol{k}}$

where the transition rate is

$$W_{1\boldsymbol{k}} = \frac{2|\gamma_{1\boldsymbol{k}}|^2}{\hbar^2} \frac{\Gamma_{1\boldsymbol{k}}}{\Gamma_{1\boldsymbol{k}}^2 + (\epsilon_{a\boldsymbol{k}} - \epsilon_{b\boldsymbol{k}})^2/\hbar^2} \qquad (14)$$

**2. Anomalous term**

Now define $Q_{\boldsymbol{k}} = < a_{\boldsymbol{k}} b_{-\boldsymbol{k}} >$

Its nonequilibrium equation is

$$\frac{d\delta Q_{\boldsymbol{k}}}{dt} = \frac{i}{\hbar} < [a_{\boldsymbol{k}} b_{-\boldsymbol{k}}, H_0 + H_2] >= -\left(\Gamma_{2\boldsymbol{k}} + i\frac{2\epsilon_{\boldsymbol{k}}}{\hbar}\right) \delta Q_{\boldsymbol{k}} - \frac{i\gamma_{2\boldsymbol{k}}}{\hbar}(\delta n_{a\boldsymbol{k}} + \delta n_{b\boldsymbol{k}})$$

The equilibrium part has already been subtracted. In steady state,

$$\delta Q_{\boldsymbol{k}} = -\frac{i\gamma_{2\boldsymbol{k}}}{\hbar} \frac{\delta n_{a\boldsymbol{k}} + n_{b\boldsymbol{k}}}{\Gamma_{2\boldsymbol{k}} + i(\epsilon_{a\boldsymbol{k}} + \epsilon_{b\boldsymbol{k}})/\hbar}$$

Again, the above quantity is related to the transition rate,

$$\frac{dn_{b-\boldsymbol{k}}}{dt} = \frac{i}{\hbar} < [b_{-\boldsymbol{k}}^{+} b_{-\boldsymbol{k}}, H_2] > = -\frac{2}{\hbar} Im(\gamma_{2\boldsymbol{k}}^{*} Q_{\boldsymbol{k}}) \equiv W_{2\boldsymbol{k}}(\delta n_{a\boldsymbol{k}} + \delta n_{b\boldsymbol{k}}) = W_{2\boldsymbol{k}} \delta n_{a\boldsymbol{k}}$$

In which

$$W_{2\boldsymbol{k}} = \frac{2|\gamma_{2\boldsymbol{k}}|^2}{\hbar} \frac{\Gamma_{2\boldsymbol{k}}}{\Gamma_{2\boldsymbol{k}}^2 + (\epsilon_{a\boldsymbol{k}} + \epsilon_{b\boldsymbol{k}})^2/\hbar^2} \qquad (15)$$

But we want the kinetic equation for **-k**, we then have,

$$\frac{dn_{b\boldsymbol{k}}}{dt} = W_{-2\boldsymbol{k}} \delta n_{a-\boldsymbol{k}}$$

For an inversion-symmetric system, $W_{2\boldsymbol{k}} = W_{-2\boldsymbol{k}}$,

and because the current distribution is odd, $\delta n_{a-\boldsymbol{k}} = -\delta n_{a\boldsymbol{k}}$

Therefore,

$$\frac{dn_{b\boldsymbol{k}}}{dt} = -W_{2\boldsymbol{k}} \delta n_{a\boldsymbol{k}}$$

This minus sign is important.

It does **not** violate momentum conservation. The anomalous process creates or annihilates and together. Pair creation preferentially stimulated by the current-carrying $a_{\boldsymbol{k}}$ distribution generates a $b_{-\boldsymbol{k}}$ excitation, which gives a counterflow in the M2 layer.

---

**3. Include relaxation in layers**

Let the ordinary intralayer relaxation in M2

$$-\delta n_{\boldsymbol{bk}}/\tau_{\boldsymbol{bk}}$$

The steady-state Boltzmann equation therefore,

$$0=-\frac{\delta n_{\boldsymbol{bk}}}{\tau_{\boldsymbol{bk}}}+(W_{1\boldsymbol{k}}-W_{2\boldsymbol{k}})\delta n_{\boldsymbol{ak}}$$

Or

$$\delta n_{\boldsymbol{bk}}=\tau_{\boldsymbol{bk}}(W_{1\boldsymbol{k}}-W_{2\boldsymbol{k}})\delta n_{\boldsymbol{ak}}$$

Therefore, the magnon current in M2 is

$$j_b^m=\frac{1}{(2\pi)^2}\int d^2k v_{\boldsymbol{k}}\tau_{\boldsymbol{bk}}[W_{1\boldsymbol{k}}-W_{2\boldsymbol{k}}]\delta n_{\boldsymbol{ak}} \qquad (16)$$

Since the transition rate from the normal dipolar coupling contains the resonant momentum transfer when $\epsilon_{\boldsymbol{ak}}=\epsilon_{\boldsymbol{bk}}$ , the abnormal dipolar coupling is usually unimportant, i.e., $W_{1\boldsymbol{k}}\gg W_{2\boldsymbol{k}}$.

## III. Magnetization configuration dependence on the quantum dipolar drag

If we assume all the other scattering parameters and the magnon band structure are isotropic, the drag current would be controlled by the angular dependence of $\gamma_{1\boldsymbol{k}}$, i.e.,

$$j_m^b\propto\int_0^{2\pi}d\phi\cos^2\phi|\gamma_{1\boldsymbol{k}}|^2$$

Note that $\cos^2\phi$in the above equation come from $v_{\boldsymbol{k}}$ and $\delta n_{\boldsymbol{ak}}$ in Eq. (16).

From Table S1, we have

| **Table S2: Configuration dependent drag current** | $\gamma_{1\mathrm{k}}/\gamma_0(k)$ | $j_m^b/j_m^b(M1=M2=+\hat{z})$ |
|---|---|---|
| Perpendicular, P: $+\hat{z},+\hat{z}$ | 1 | 1 |
| Perpendicular, AP: $+\hat{z},-\hat{z}$ | $e^{2i\phi}$ | 1 |
| In-plane, P: $+\hat{x},+\hat{x}$ | $-\cos^2\phi$ | 5/8 |
| In-plane, AP: $+\hat{x},-\hat{x}$ | $(1-\sin\phi)^2$ | 21/8 |
| In-plane, P: $+\hat{y},+\hat{y}$ | $-\sin^2\phi$ | 1/8 |
| In-plane, AP: $+\hat{y},-\hat{y}$ | $(1-\cos\phi)^2$ | 49/8 |